\documentclass[a4paper,twoside]{article}

\usepackage[percent]{overpic}
\usepackage{epsfig}
\usepackage{subcaption}
\usepackage{calc}
\usepackage{amssymb}
\usepackage{amstext}
\usepackage{amsmath}
\usepackage{amsthm}
\usepackage{multicol}
\usepackage{multirow}
\usepackage{pslatex}
\usepackage{apalike}
\usepackage{url}
\usepackage{color}
\usepackage{SCITEPRESS}     

\usepackage{microtype}

\begin{document}

\title{Multi-view data capture using edge-synchronised mobiles}

\author{\authorname{Matteo Bortolon, Paul Chippendale, Stefano Messelodi and Fabio Poiesi}
\affiliation{Fondazione Bruno Kessler, Trento, Italy}
\email{matteo.bortolon@studenti.unitn.it, \{chippendale, messelod, poiesi\}@fbk.eu}
}

\keywords{Synchronisation, Free-Viewpoint Video, Edge Computing, Augmented Reality, ARCloud.}

\abstract{
Multi-view data capture permits free-viewpoint video (FVV) content creation.
To this end, several users must capture video streams, calibrated in both time and pose, framing the same object/scene, from different viewpoints.
New-generation network architectures (e.g.~5G) promise lower latency and larger bandwidth connections supported by powerful edge computing, properties that seem ideal for reliable FVV capture.
We have explored this possibility, aiming to remove the need for bespoke synchronisation hardware when capturing a scene from multiple viewpoints, making it possible through off-the-shelf mobiles.
We propose a novel and scalable data capture architecture that exploits edge resources to synchronise and harvest frame captures.
We have designed an edge computing unit that supervises the relaying of timing triggers to and from multiple mobiles, in addition to synchronising frame harvesting.
We empirically show the benefits of our edge computing unit by analysing latencies and show the quality of 3D reconstruction outputs against an alternative and popular centralised solution based on Unity3D.}

\onecolumn \maketitle \normalsize \setcounter{footnote}{0} \vfill

\section{Introduction}\label{sec:introduction}
Immersive computing represents the next step in human interactions, with digital content that looks and feels as if it is physically in the same room as you. 
The potential for immersive digital content impacts upon entertainment, advertising, gaming, mobile tele-presence and tourism \cite{Shi2015,Jiang2017,Elbamby2018,Rematas2018,Park2018,Qiao2019}.
But for immersive computing to become mainstream, a vast amount of easy-to-create free-viewpoint video (FVV) content will be needed. 
The production of 3D digital objects inside a real-world space is almost considered a solved problem \cite{Schonberger2016}, but doing the same for FVV is still a challenge \cite{Richardt2016}. 
This mainly because FVVs require the temporal capturing of dynamic objects from different, and calibrated viewpoints \cite{Guillemaut2011,Mustafa2017}. 
Synchronisation across viewpoints dramatically reduces reconstruction artefacts in FVVs \cite{Vo2016}. 
For controlled setups, frame-level synchronisation can be achieved using shutter-synchronised cameras \cite{Mustafa2017}, however this is impractical in uncontrolled environments with conventional mobiles. 
Moreover, mobile pose estimation (i.e.~position and orientation) with respect to a global coordinate system is necessary for content integration.
This can be achieved using frame-by-frame Structure from Motion (SfM) \cite{Schonberger2016} or Simultaneous Localisation And Mapping (SLAM) \cite{Mur-Artal2015}. 
The latter has shown to be effective for collaborative Augmented Reality though the ARCloud \cite{anchors}.
ARCore Anchors\footnote{The AR Anchor is a rigid transformation from local to a global coordinate system.} are targeted at AR multiplayer gaming, but we experimentally tested that they are not yet ready for FVV production as is, i.e.~relative pose estimation is not sufficiently accurate for 3D reconstruction.
As per our knowledge a flexible and scalable solution for synchronised calibrated data capture deployable on conventional mobiles does yet not exist.
\begin{figure*}[t]
\includegraphics[width=\textwidth]{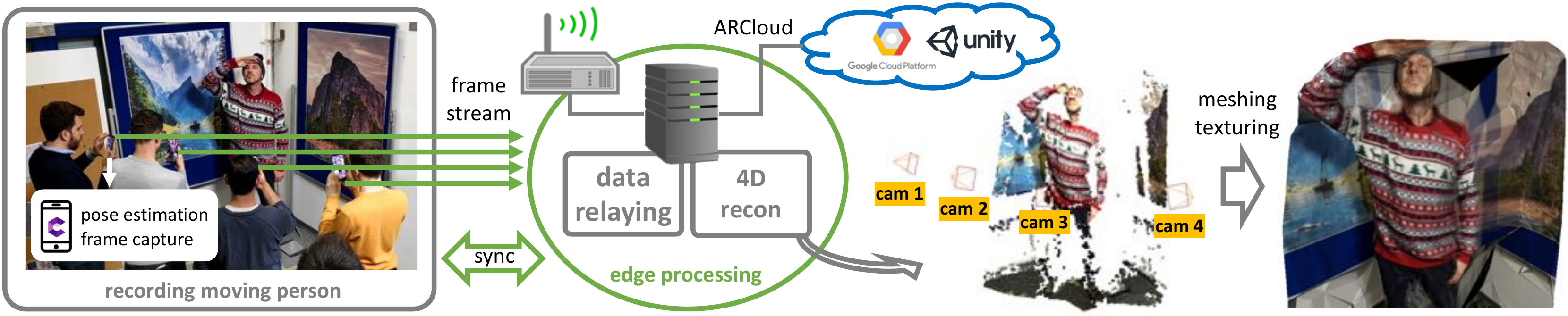}
\caption{Multi-view data captures for free-viewpoint video creation generates large data throughput, and requires synchronised and calibrated cameras.
Our solution offloads computations from mobiles and the cloud to the edge, handling synchronisation and image processing more efficiently.
Moving processing closer to the user improves performance and fosters scalability.}
\label{fig:teaser}
\end{figure*}
As FVVs require huge data transmissions, throughput is another challenge for immersive computing.
Fig.~\ref{fig:teaser} shows four people recording a person with their mobiles. 
The measured throughput in this setup was about 52Mbps with frames captured at 10Hz, with a resolution of 640$\times$480 pixels and encoded in JPEG. 
Note that only a portion of the recorded person was covered. 
For a more complete, 360-degree coverage, several more mobiles would be needed. 
With the introduction of new wireless architectures that target high-data throughput and low latency, such as Multi-access Edge Computing (MEC), device-to-device communications and network slicing in 5G networks, scalable communication mechanisms that are appropriate for the deployment of immersive content on consumer devices are key \cite{Shi2016,Qiao2019}. 
Hybrid cloud/fog/edge solutions will ensure that users get low-lag feedback as well as the possibility to offload computationally intensive tasks.

In this paper, we present a system that harvests data generated from multiple-handheld mobiles at the edge instead of harvesting it in the cloud.
This promotes scalability, lower latency and facilitates synchronisation.
Our implementation consists of a server, namely Edge Relay, that handles communications across mobiles and a Data Manager that harvests the content captured by the mobiles.
We handle synchronisation through a relay server because, as opposed to a peer-to-peer one, relay servers can effectively reduce bandwidth usage and improve connection quality \cite{Hu2016}.
Although relay servers may lead to increases of network latency, peer-to-peer connection may be ineffective because when mobiles are within different networks (e.g.~different operators in 5G networks), they cannot retrieve their respective IP address due to Network Address Translation (NAT).
Moreover, to mitigate the latency problem, we designed a latency compensation strategy, that we empirically tested to be effective when the network conditions are fairly stable.
We developed an app that each mobile uses to capture frames and estimate pose with respect to a global coordinate system through ARCore \cite{arcore}. 
As the relative poses from ARCore are not accurate enough for FVV, we refine them using SfM \cite{Schonberger2016}.
To summarise, the key contributions of our system are (i) the protocols we have designed to allow users under different networks (or operators) to join the same data capture session, (ii) the integration of a latency compensation mechanism to mitigate the communication delay among devices, and (iii) the integration of these modules with a SLAM framework to estimate mobile pose in real time and to globally localise mobiles in an environment through the ARCloud.
As per our knowledge, this is the first proof-of-concept, decentralised system for synchronised multi-view data capture usable for FVV content creation.
We have carried out a thorough experimental analysis by jointly assessing latency and temporal 3D reconstruction.
We have compared results against an alternative and popular centralised solution based on Unity3D \cite{unity_relay_server}.

\section{Related work}

\noindent\textbf{Low Latency Immersive Computing:} To foster immersive interactions between multiple users in augmented spaces, low-latency computing and communications must be supported \cite{Yahyavi2013}. Although mobiles are the ideal medium to deliver immersive experiences, they have finite resources for complex visual scene understanding, reasoning and graphical tasks, hence computational offloading is preferred for demanding and low-latency interactions. Fog and edge computing, soon to be mainstream thanks to 5G, will be one of the key enablers. Thankfully, not all immersive computing tasks (e.g.~scene understanding, gesture recognition, volumetric reconstruction, illumination estimation, occlusion reasoning, rendering, collaborative interaction sharing) have the same time-critical nature. \cite{Chen2016,Zhang2018} showed that scene understanding via object classification could be performed at a rate of several times per minute by outsourcing computations to the cloud. \cite{Zhang2018} showed how an optimised image retrieval pipeline for a mobile AR application can be created by exploiting fog computing, reducing data transfer latency up to five times compared to cloud computing. \cite{Sukhmani2018} analysed the concept of dynamic content caching for mobiles, i.e.~what to cache and where, and they illustrated that a dramatic performance increase could be obtained by devising appropriate task offloading strategies. \cite{Bastug2017} showed how a pro-active content request strategy could effectively be used to predict content before it was actually requested, thus reducing immersive experience latency, at the cost of increased data overhead.
In the cases of FVV, which is very sensitive to synchronisation issues, communications must be executed as close to the user as possible to reduce lag. Solutions to address the problem can be hardware or software based. Hardware-based solutions include timecode synchronisation with or without genlock \cite{Kim2012,Wu2008}, and Wireless Precision Time Protocol \cite{Garg2018}. Hardware based solutions is not our target as they require important modifications to the communication infrastructure. Software-based solutions are often based on the Network Time Protocol (NTP), instructing devices in a session to acquire frames at prearranged time intervals and then attempt to compensate/anticipate for delays \cite{Latimer2015}. Cameras can share timers that are updated by a host camera \cite{Wang2015}. Alternatively, errors in temporal frame alignment have been addressed using spatio-temporal bundle adjustment, in an offline post-processing phase \cite{Vo2016}. However, this type of post-alignment also incurs a high computational overhead, as well as adding more latency to the creation and consumption of FVV reconstructions. Although NTP approaches are simple to implement, they are unaware of situational-context. Hence, the way in which clients are instructed to capture images in a session is totally disconnected from scene activity, hence they are unable to optimise acquisition rates either locally or globally, prohibiting optimisation techniques such as \cite{Poiesi2017}, that aim to save bandwidth and maximise output quality. Our solution operates online and is aimed at decentralising synchronisation supervision, thus is more appropriate for resource-efficient, dynamic-scene capture.

\noindent\textbf{Free-viewpoint video production:} Free-viewpoint (volumetric or 4D) videos can be created either through the synchronised capturing of objects from different viewpoints \cite{Guillemaut2011,Mustafa2017} or with Convolutional Neural Networks (CNN) \cite{Rematas2018} that estimate unseen content. The former strategy needs camera poses to be estimated/known for each frame, using approaches like SLAM \cite{Zou2013} or by having hardware calibrated camera networks \cite{Mustafa2017}. Typically, estimated poses lead to less-accurate reconstructions \cite{Richardt2016}, when compared to calibrated setups \cite{Mustafa2017}. Converserly, CNN-based strategies do not need camera poses, but instead need synthetic training data of 3D objects extracted, for example, from video games or Youtube videos \cite{Rematas2018} as they need to estimate unobserved data.
Traditional FVV (i.e.~non-CNN) approaches can be based on shape-from-silhouette (SFS) \cite{Guillemaut2011}, shape-from-photoconsistency (SFP) \cite{Slabaugh2001}, multi-view stereo (MVS) \cite{Richardt2016} or deformable models (DM) \cite{Huang2014}. SFS methods aim to create 3D volumes (or visual hulls) from the intersections of visual cones formed by 2D outlines (silhouettes) of objects visible from multiple views. SFP methods create volumes by assigning intensity values to voxels (or volumetric pixels) based on pixel-colour consistencies across images. MVS methods create dense point clouds by merging the results of multiple depth-maps computed from multiple views. DM methods try to fit known reference 3D-models to visual observations, e.g.~2D silhouettes or 3D point clouds. All these methods need frame-level synchronised cameras. \cite{Vo2016} proposed a spatio-temporal bundle adjustment algorithm to jointly calibrate and synchronise cameras. Because it is a computationally costly algorithm, it is desirable to initialise it with ``good" initial camera poses and synchronised frames. Amongst these methods, MVS produces reconstructions that are geometrically more accurate than the other alternatives, albeit at a higher computational cost. Approaches like SFS and SFP are more suitable for online applications as they are fast, but outputs have less definition.

\section{Data capturing overview}\label{sec:volumetric_capturing}
\begin{figure}[t]
\begin{center}
\includegraphics[width=1\columnwidth]{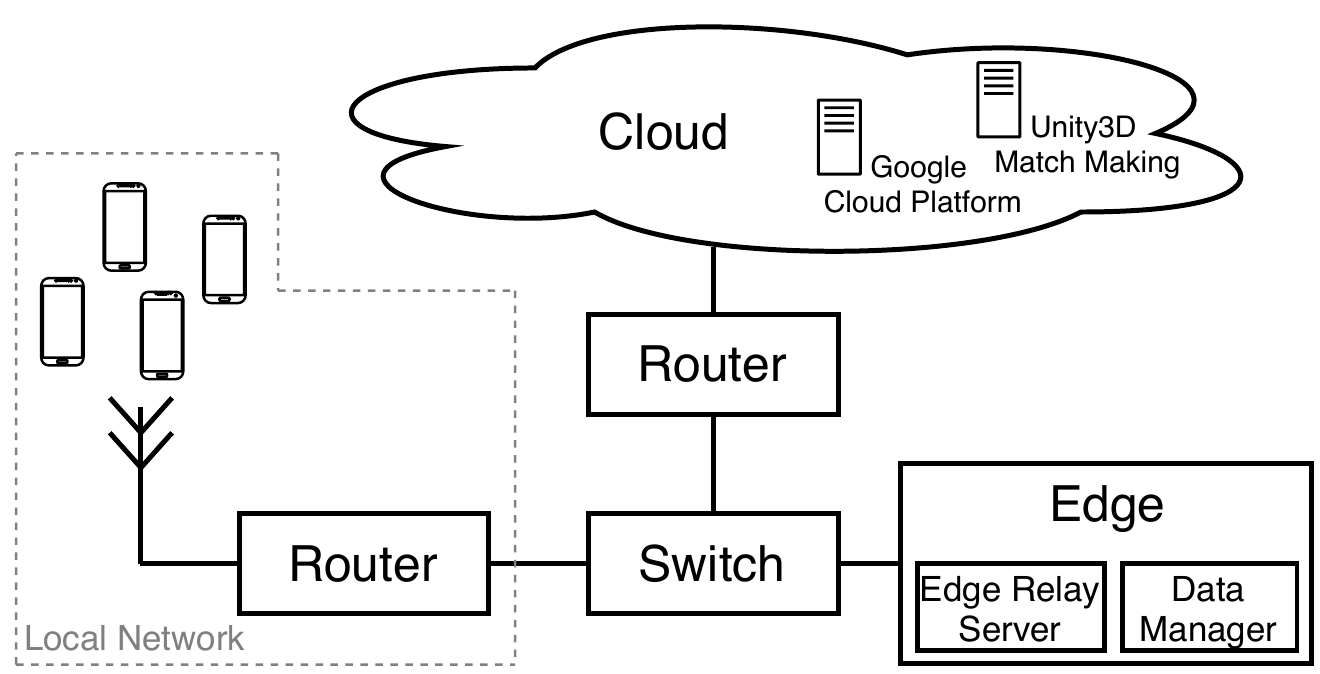}
\end{center}
\vspace{-.2cm}
\caption{Block diagram of our edge-based architecture. Mobiles are connected to the same local network (e.g.~WiFi or 5G). When they perform a FVV capture, data relaying and processing is performed on the edge, ensuring low-latency and synchronised frame capture.}
\label{fig:diagram}
\end{figure}

\begin{figure}[t]
\begin{center}
\includegraphics[width=.9\columnwidth]{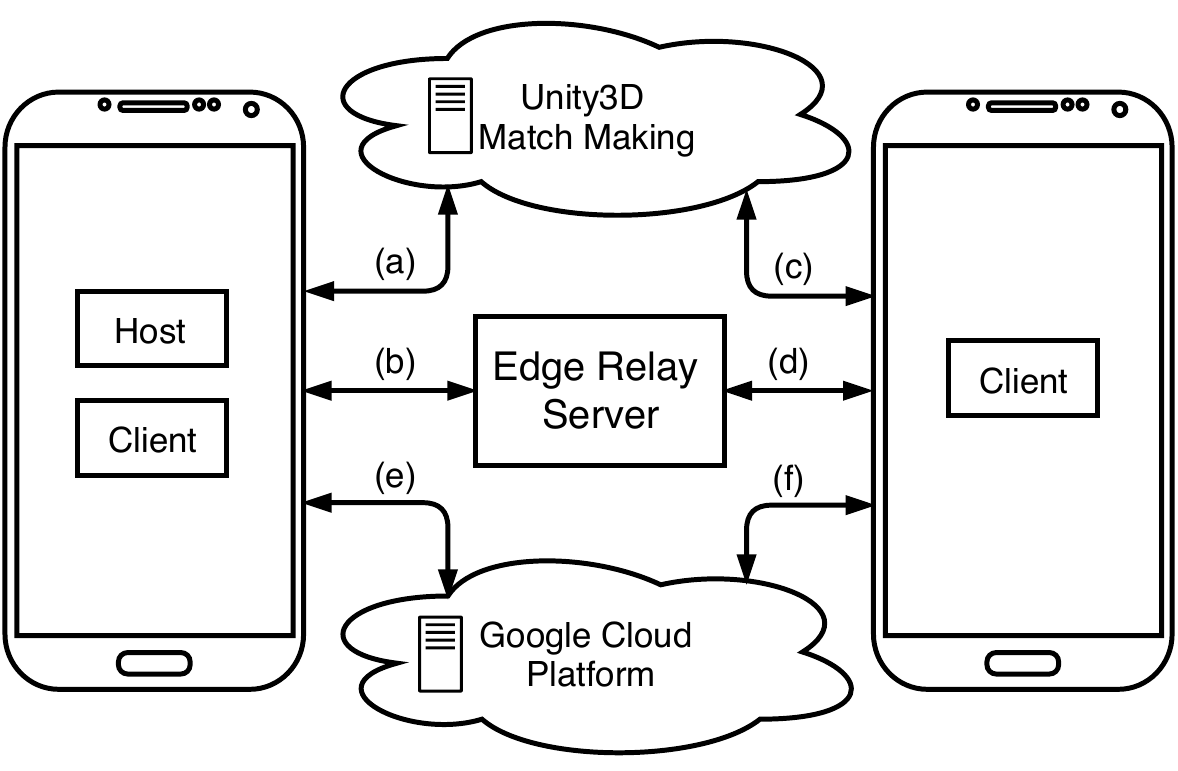}
\end{center}
\vspace{-.2cm}
\caption{Session setup procedure. Description of the steps can be found in text.}
\label{fig:ue_comm}
\end{figure}

Our system carries out synchronisation and data capture at the edge, and uses cloud services for the session initialisation.
The edge hosts two applications: an Edge Relay and a Data Manager.
Services used in the cloud are a Unity3D Match Making server and the Google Cloud Platform.
Fig.~\ref{fig:diagram} shows the block diagram of our system. Fig.~\ref{fig:ue_comm} details the session setup.

When a group of users want to initiate a multi-view capture, they must connect their mobiles to a wireless network via WiFi or 5G, and then open our frame-capture app.
One user, designated the session host, creates a new session in the app. 
This sends a request to the Match Making server \cite{matchmaking} to create an acquisition session (a).
This request includes the IP address of the host as seen from the Edge Relay.
The Match Making server adds this session to a list of active sessions, associating it with a unique session ID.
The Match Making server publicises this session ID for others to join as clients.
The host device informs the Edge Relay that it is ready to accept client devices (b). 
Client devices see the list of active sessions from the Match Making server and they choose one to join (c).
When a client chooses a session, they retrieve the session ID, the host's IP address from the Match Making server, and it uses this information to connect to the Edge Relay (d).
The Edge Relay validates client connections by verifying that the information provided is correct.
When all clients have joined a session, the Match Making server is not used again.

Before starting the frame capture, (i) all users must map their local surroundings in 3D using the app's built-in ARCore functionality, then (ii) the host measures the communication latency between itself and the clients to inform the clients of the compensation needed to handle network delays.
Latency compensation is explained in Sec.~\ref{sec:Edge_Relay}.
3D mapping involves capturing sparse geometric features of the environment \cite{Cadena2016}.
Once mapping is complete, the host user places an AR Anchor in the mapped scene to define a common coordinate system for all devices (e).
This AR Anchor is uploaded by the host to the Google Cloud Platform and then automatically downloaded by clients through HTTPS \cite{RFC7540}, or the QUIC protocol by devices that support it \cite{quic} (f). 
Finally, the capture session starts when the host client presses `start' on their app.

Snapshots, or captured frames, can be taken either periodically \cite{Knapitsch2017}, or dynamically based on scene content \cite{Resch2015,Poiesi2017}. 
Frame captures based on scene content is desirable because one can avoid excessive data traffic when a scene is still and then capture fast dynamics by increasing the rate when high activity is observed. 
However, the latter is more challenging than the former as it requires mobiles to perform on-board processing and a decentralised mechanism to reliably relay synchronisation signals.
We designed our system to be suitable for dynamic frame captures, hence we have chosen to let the host mobile drive synchronisation, rather than fixing the rate beforehand on an edge or cloud server.
To trigger the other mobiles to capture a frame, the host sends a snapshot (or synchronisation) \emph{trigger} to the Edge Relay. 
Snapshot triggers instruct mobiles in a session to capture frames.
The Edge Relay forwards triggers received, from the host, to all clients, instructing them to capture frames and generate associated meta-data (e.g.~mobile pose, camera parameters).
Captured frames and meta-data are momentarily buffered on the devices and then transmitted to the Data Manager asynchronously.
Without loss of generality, the host uses an internal timer to take snapshots that expire every $C = 1/F$. 
A \emph{snapshot counter} is incremented each time the countdown expires and it is used as unique identifier for each snapshot taken.

\section{Edge Relay}\label{sec:Edge_Relay}
Traditional architectures for creating multi-user experiences are based on authoritative servers \cite{Yahyavi2013}, typically exploiting relay servers \cite{unity_relay_server,photon}.
An authoritative-server based system allows one of the participants to be both a client and the host at the same time, thus having the authority to manage the session \cite{hlapi}.
Our Edge Relay routes session control messages from the host to the clients via UDP, to avoid delays caused by flow-control systems (e.g.~TCP).
The Edge Relay handles four different types of messages: \emph{Start Relaying}, \emph{Connect-to-Server}, \emph{Data} and \emph{Disconnect Client}.

The host makes a \emph{Start Relaying} request to the Edge Relay to begin a session.
This request carries the connection configuration (e.g.~client disconnection deadline, maximum packet size, host's IP address and port number), which is used as a verification mechanism for all the clients to connect to the Edge Relay \cite{mlapi_config}.
The verification is performed through a cyclic redundancy check (CRC).
If the host is behind NAT, clients will not be able to retrieve its IP address nor the port number, hence they will not be able to include them in the connection configuration, and hence they will not pass the verification stage.
In order to mitigate this NAT-related issue, we required the Edge Relay to communicate to the host the IP address and port number with which the Edge Relay sees the host.
This IP address can be the actual IP of the host if the Edge Relay and host are within the same local network, or the IP address of the router (NAT) if host and Edge Relay are on different networks.
After the host receives this information from the Edge Relay, the host communicates host's IP address and port number to the Match Making server.
In this way, when the clients discover the session ID from the Match Making server, they can retrieve the host's IP address and port number, and use them for verification to connect to the Edge Relay.

When a client decides to join a session, the client sends a \emph{Connect-to-Server} request message to the Edge Relay. 
This message contains the IP and port address of the host, which the client retrieved from the Match Making server. 
The Edge Relay checks to see if the requested session associated to this IP address and port is already hosted by a mobile.
If it is, then the Edge Relay adds this client to the list of session participants.

\emph{Data} messages carry information from one device to another in a session. When a data packet is received by the Edge Relay it explores the header to understand where the packet must be forwarded to: either to specific devices or broadcast to all. We use a data messaging system that involves two types of messages: State Update packages or Remote Procedure Calls (RPCs) \cite{data_messages}. 
State Update packages are used to update elements in the session and to propagate the information to all participants, i.e.~the AR Anchor, while RPCs are used for control commands, i.e.~synchronisation triggers and the latency estimation mechanism. 
The RPC of the synchronisation trigger carries the information of the snapshot counter (Sec.~\ref{sec:volumetric_capturing}).

A \emph{Disconnect Client} message is exchanged when a user exits the session. 
This message can be sent by the client or by the Edge Relay. 
The Edge Relay detects the exit of a client if a \emph{Keep-alive} packet is not received within a timeout. 
We set the Keep-alive time at 100ms and disconnect timeout as 1.6s. 
Upon disconnection of a client, the Edge Relay informs all the other participants and removes this device from the list of participants of the session.

\section{Frame capture app optimisations}\label{sec:frame_capture}
To deal with latency variation and large throughput, we have implemented two optimisation strategies.

The latency between client/host and the Edge Relay can vary due to the distance between devices and the antenna, network traffic, or interference with other networks \cite{Soret2014}.
A high latency can negatively affect the geometric quality of the reconstructed object, so it must be understood and compensated for \cite{Vo2016}. 
To cope with network latency issues, we have implemented a latency compensation mechanism that uses Round Trip Time measurements on the communication link between host and clients. 
We model the latency measured between devices to delay the capture of a frame upon the reception of a synchronisation trigger for each device independently.
This enables the devices to capture frames (nearly) synchronously.
Specifically, during the initialisation phase, the host builds a $N \times M$ matrix $P$, where the element $p(i,j)$ is the $j$-th measured Round Trip Time (i.e.~ping) between the host and the $i$-th client. $N$ is the number of clients and $M$ is the total number of ping measurements. The host computes the average ping for each client, such as $\bar{p}(i) = \frac{1}{M} \sum_{j=1}^M p(i,j)$, and extracts the maximum ping as $\hat{p} = max(\{\bar{p}(1),...,\bar{p}(N)\})$. Then the $i$-th client captures the snapshot with a delay of $\Delta t_i = \frac{1}{2} \cdot (\hat{p} - \bar{p}(i))$ ms, whereas the host captures the snapshot with a delay of $\hat{p}$ ms.

Each time a client receives a synchronisation trigger, it captures a frame, and the associated meta-data, i.e.~pose (with respect to the global coordinate system), camera intrinsic parameters (i.e.~focal length, principal point), device identifier and snapshot counter.
To effectively handle frame captures when synchronisation triggers are received, we use two threads on the mobiles. 
Triggers are managed by the main thread, which uses a scheduler to guarantee that a frame is captured when the calculated delay $\Delta t_i$ expires. 
Each captured frame is passed to a second thread that encodes it into a chosen format (e.g.~JPEG) and enqueues it for transmission to the Data Manager. 
If there is bandwidth available over the communication channel, frames are transmitted immediately, otherwise they are buffered. 
In the next section we explain how the Data Manager processes the received frames.

\section{Data Manager}\label{sec:Data_Manager}

\begin{figure}[t]
\begin{center}
\includegraphics[width=\columnwidth]{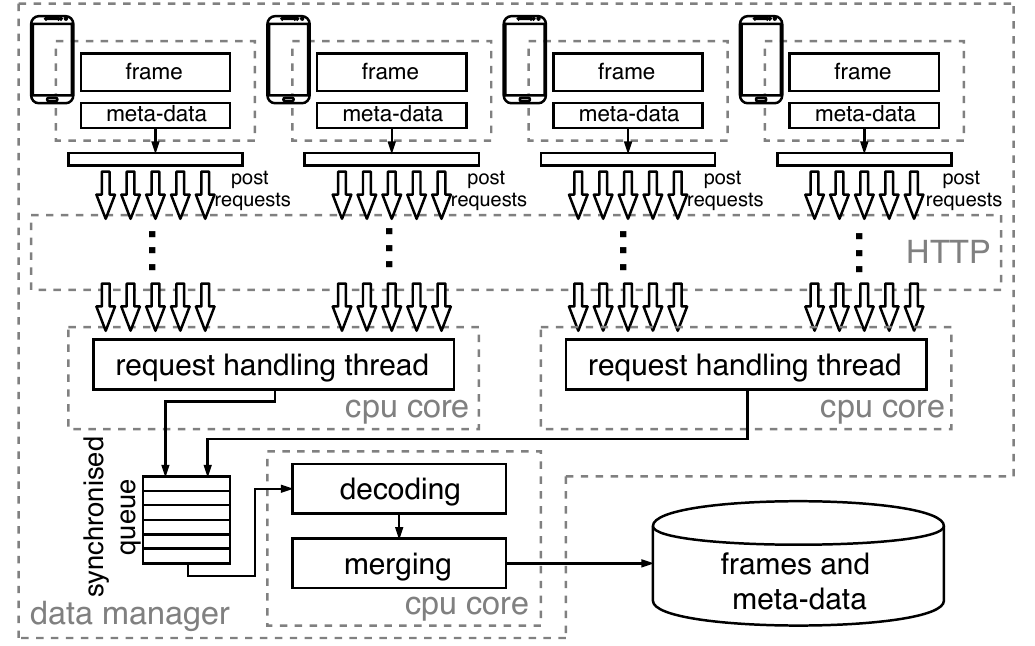}
\end{center}
\vspace{-.2cm}
\caption{Data manager architecture.}
\label{fig:data_manager}
\end{figure}

The Data Manager is an application that resides on the edge unit and functions independently from the Edge Relay (Fig.~\ref{fig:diagram}). 
The communication between the Data Manager and the participants is based on HTTP requests \cite{RFC1945}. 
Fig.~\ref{fig:data_manager} shows the architecture of our Data Manager.
The Data Manager's operation consists of three phases: \emph{Stream Initialisation}, \emph{Frame Transmission} and \emph{Stream End}.

During the \emph{Stream Initialisation} phase, the host sends an initial HTTP request to the Data Manager containing information about the number of devices that it should expect to receive data from within a session. 
When this request is received, the Data Manager creates a unique Stream ID for the session, which is sent to the host and clients. Then, the Data Manager initialises a new thread to perform decoding and merging operations (explained later).

\emph{Frame Transmission} occurs when the criteria for taking a snapshot has been met (Sec.~\ref{sec:frame_capture}). In particular, a client, after it receives the RPC and after it compensates for the latency, sends the requested frame along with the meta-data to the Data Manager through a HTTP request. We measured that the Data Manager can process a HTTP request in about 200 to 300 ms. To optimise the HTTP-request ingestion rate on the Data Manager, each mobile creates multiple and simultaneous HTTP requests that will be processed in parallel by Request Handling Threads. We create as many Request Handling Threads as the number of CPU cores available.
Then, we measured that a mobile can process up to 12 simultaneous requests with negligible computational time and that the Data Manager can handle up to 100 requests. Therefore we create a policy where, if $N$ is the number of mobiles connected, each mobile can create up to $r = min\{\lfloor N/100 \rfloor, 12\}$ HTTP requests, where $\lfloor \cdot \rfloor$ is the rounding to lower integer operation. Each Request Handling Thread processes each HTTP request and pushes it into a synchronised queue, which in turn feeds a decoder in charge of converting frames into a single format (e.g.~in JPG, PNG). We measured that this operation can handle up to four mobiles transmitting at 20 fps in real-time. Lastly, we use a merging operation to re-organise data based on their snapshot counter. If the merging operation detects that, for a given snapshot trigger, the number of frames received is not the same as the number of mobiles connected, the received frames will be labelled as \emph{partial} when stored in the database, so the FVV reconstruction algorithm can handle them accordingly.

A \emph{Stream End} occurs when the host ends a capture session. In addition to stopping frame acquisition, the host also sends a request to terminate the session to the Data Manager, which in turn waits until the last acquired snapshots have been received before terminating of the opened Threads.

\section{Results}\label{sec:results}

\noindent\textbf{Motivation:}
Evaluating multi-view data capture quantitatively is challenging because both pose estimation and synchronisation should be assessed.
A possibility could be to create a FVV using the captured frames and assess the output quality.
However, FVV ground truth is difficult to obtain, especially when an object being reconstructed is non-rigid.
\cite{Mustafa2017,Richardt2016} mainly evaluated their FVV outputs qualitatively, and selected sub-modules for the quantitative assessment.
Based on a similar idea, we have performed a qualitative analysis consisting of a live recording using handheld mobiles.
We used four mobiles (two Huawei P20Pro, one OnePlus Five and one Samsung S9) simultaneously observing a moving person and we reconstructed this person using a popular SfM technique, i.e.~COLMAP \cite{Schonberger2016}.
Then, we quantified the performance of our system under controlled conditions by evaluating the reconstruction accuracy (3D triangulation error) of a rotating texture-friendly object.
Lastly, to explicitly determine the end-to-end time difference of the acquired frames we performed a two-view frame capture of a stopwatch displayed on an iPad screen.

\noindent\textbf{Implementation:} 
Our Edge Relay is based on MLAPI \cite{mlapi} and is developed in C\#. 
The Data Manager is developed in Python. 
Both applications are run in a Docker container to facilitate deployment. 
Our Edge Relay is a laptop with CPU i7 and 16GB RAM. 
The application running on mobiles is developed in Unity3D using ARCore 1.7 and OpenCV 4.0 libraries.
Captured frames have a size of 640$\times$480 pixels.

\subsection{Experimental setup}

\begin{figure}[t]
\begin{center}
	\begin{tabular}{@{}c@{}c}
		\includegraphics[width=.5\columnwidth]{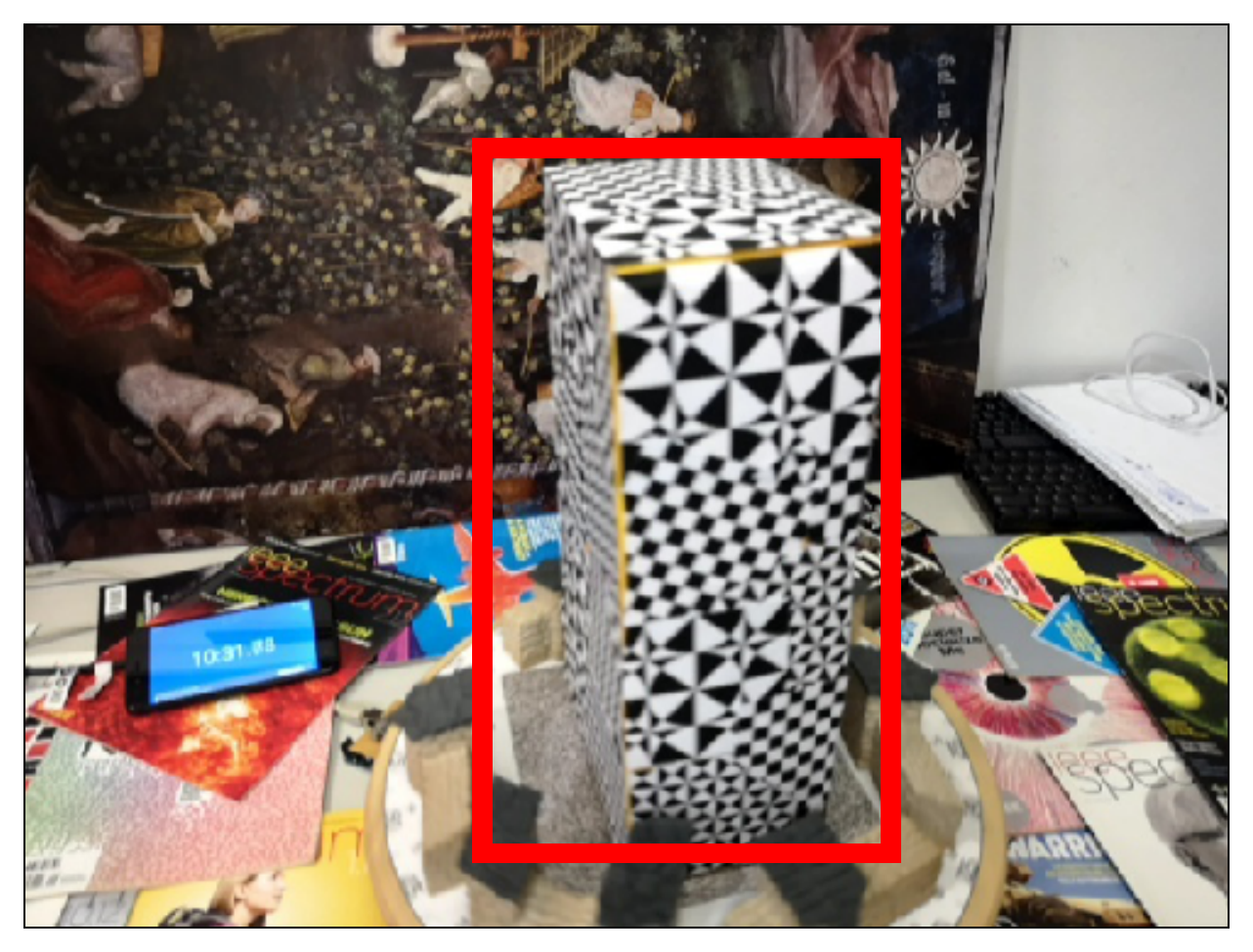}&\includegraphics[width=.5\columnwidth]{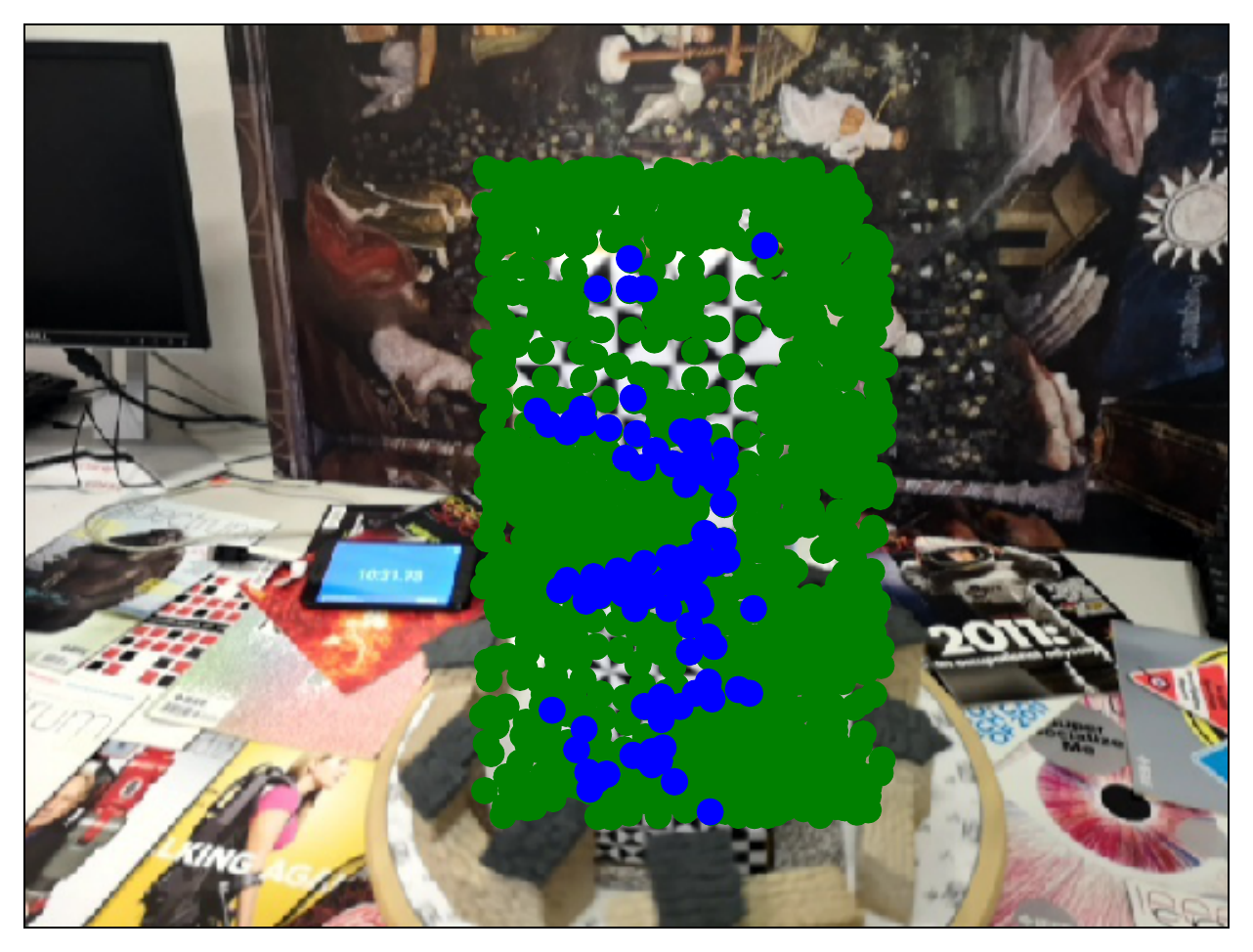}\\
        (a)&(b)\\
	\end{tabular}
\end{center}
\vspace{-.2cm}
\caption{Experimental setup: (a) Left-hand camera: red box shows region of interest for our analysis. (b) Right-hand camera: green points show keypoints extracted and blue points show keypoints that have been 3D triangulated.}
\label{fig:exp_setup}
\end{figure}

\noindent\textbf{3D Reconstruction assessment:}
We placed a reference object on an adjustable angular-velocity turntable, and rotated it 270 degrees clockwise and then 270 degrees anticlockwise. 
We 3D-reconstructed the reference object over time from images captured from two Huawei P20Pro positioned on tripods: vertically at the same height, horizontally at 20cm far from each other, and 40cm far from the rotating object.
We used tripod mounts to reduce pose-estimation errors, as we wanted to quantify reconstruction errors brought about by network lag and synchronisation effects.
Pairs of frames with the same snapshot counter are fed into COLMAP.
The 3D-reconstruction algorithm processed all the pairs captured in the experiment.
Fig.~\ref{fig:exp_setup}a shows the object from the left-hand camera; the red bounding box highlights the region of interest we have used for our analysis of keypoints/3D points. 
Fig.~\ref{fig:exp_setup}b shows the view from the right-hand camera with the keypoints, highlighted in green, and the keypoints that have been 3D triangulated, shown in blue.

Latencies have been compared to those obtained using the Unity3D Relay \cite{unity_relay_server}.
We also assessed the quality of 3D reconstructions over time by comparing the volumetric models of the reference object under different network latencies with respect to a ground truth, which was created by reconstructing the reference object in 3D, frame-by-frame with one degree of separation between frames, for a total of 270 degrees. 
This created two sequences of 270 aligned frames, one sequence for each camera. 
By picking one frame from the first sequence and then another frame from the second sequence captured with a different pose, we could simulate different angular velocities and different frame-capture rates.
For example, say we wanted to simulate the reference object rotating at $50deg/s$, captured at $F=10Hz$. 
This corresponds to a 100ms interval between frames, corresponding to an object rotation of $5deg$. 
We can pick frame $t$ from the first sequence and frame $t+5$ from the second sequence to simulate this condition.
In our experiments we simulated angular velocities between $50 deg/s$ and $100 deg/s$ with steps of $10 deg/s$, and modelled snapshots that were captured with a frequency of $F=10Hz$. 
We then modelled the latency between the mobile and the Edge Relay by adding normally distributed delays, i.e.~$\mathcal{N}(\mu, \sigma)$, where $\mu$ is the mean and $\sigma$ is the standard deviation that we measured on our experimental WiFi network. 
In order to use realistic latency estimates, we recreated latency variation conditions, ranging from $0$ms to $150$ms with a step of $30$ms, by injecting delays into the network using NetEm \cite{NetEm}. 
The mobiles were connected via a WiFi 2.4GHz network. 
We used an off-the-shelf WiFi access point (Thomson TG587nV2). 
Due to interference caused by neighbouring networks (a typical scenario nowadays), we observed typical Round Trip Times (RTT) shown in Tab.~\ref{tab:mobile2edge}.
\begin{table}[t]
\caption{Round Trip Time between mobile and Edge Relay measured on our WiFi network. `Set' are the latencies set with NetEm \cite{NetEm}, and `Meas' are mean $\pm$ standard deviation calculated over 40 Round Trip time measurements.}
\label{tab:mobile2edge}
\tabcolsep 2pt
\resizebox{\columnwidth}{!}{%
\begin{tabular}{l|c|c|c|c|c|c}
Set (ms) & 0 & 30 & 60 & 90 & 120 & 150 \\
\hline
Meas (ms) & $14\pm11$ & $50\pm14$ & $80\pm31$ & $108\pm26$ & $141\pm26$ & $167\pm15$ \\
\end{tabular}
}
\end{table}

We used the data in Tab.~\ref{tab:mobile2edge} to quantify the triangulation error between ground-truth 3D points and 3D points triangulated under simulated delays. We calculated the triangulation error using a grid composed of 16$\times$16-pixel cells defined within the bounding box shown in Fig.~\ref{fig:exp_setup}a. Within each cell we select the keypoints that have been triangulated in 3D and calculated the centre of mass of their 3D projection. We then calculated the Euclidean distance between the 3D centres of mass of the ground truth and the 3D centres of mass of the points triangulated with different latencies.

\noindent\textbf{End-to-end delay assessment:} 
We used two mobiles configured as the 3D reconstruction case to capture the time ticked by a stopwatch (up to the millisecond precision).
We extracted the time information from each frame pair of frames using OCR \cite{ocr} and computed their time difference.
We performed this experiment using the delay compensation activated.
The configurations tested are with our Edge Relay operating locally (i.e.~at the edge) and in the cloud.
For the latter we deployed our Edge Relay on Amazon Web Services Cloud (AWS), and connected the mobiles to the cloud through a high-quality optical-fibre-based internet connection and though 4G, to resemble a real capture scenario.

\subsection{Experiments}

\noindent\textbf{Relay latency:}
We assessed synchronisation trigger delays by measuring the latency between the host and clients in the case of our Edge Relay and the Unity3D Relay. 
We performed these measurements with a dedicated feature integrated into the app that produces 250 RTT measurements. We then calculated the average measured RTTs.
Unity3D Relay is cloud based and can be located anywhere around the globe, based on a user's location, the closest is usually queried \cite{unity_match_host}.
In the case of our Edge Relay, we measured the host-client RTT and obtained an average of $66\pm37$ms.
Then we measured the mobile-Edge Relay RTT and obtained an average of $14\pm11$ms.
This means that $66 - 14\cdot2 = 30$ms is consumed by the network devices (e.g.~router, access point) and by the Edge Relay for processing.
In the case of Unity3D Relay \cite{unity_relay_server}, we measured the host-client RTT and obtained an average of $89 \pm 12$ms.
Then we measured the mobile-Unity3D Relay RTT and obtained an average of $28 \pm 13$ms.
This means that $89 - 28\cdot2 = 33$ms is consumed by network devices to reach the cloud and by the Unity3D Relay for processing.
The Unity3D Relay's processing time is comparable to that of our Edge Relay, but more stable, as the standard deviation is smaller. 
We believe that we can increase the efficiency of our Edge Relay by re-implementing some core modules of MLAPI \cite{mlapi} in C++.
Furthermore, the RTT of $28$ms in the case of Unity3D Relay has been measured inside a research centre with a high-quality optical-fibre-based internet connection.
To have an idea of how other types of internet connections could affect latency, we performed the same RTT measurement towards Unity3D Relay but using a traditional 6Mbs home broadband and a 4G connection, which turned out to be $255 \pm 66$ms and $80 \pm 7$ms, respectively.

\begin{figure}[t]
\begin{center}
\includegraphics[width=1\columnwidth]{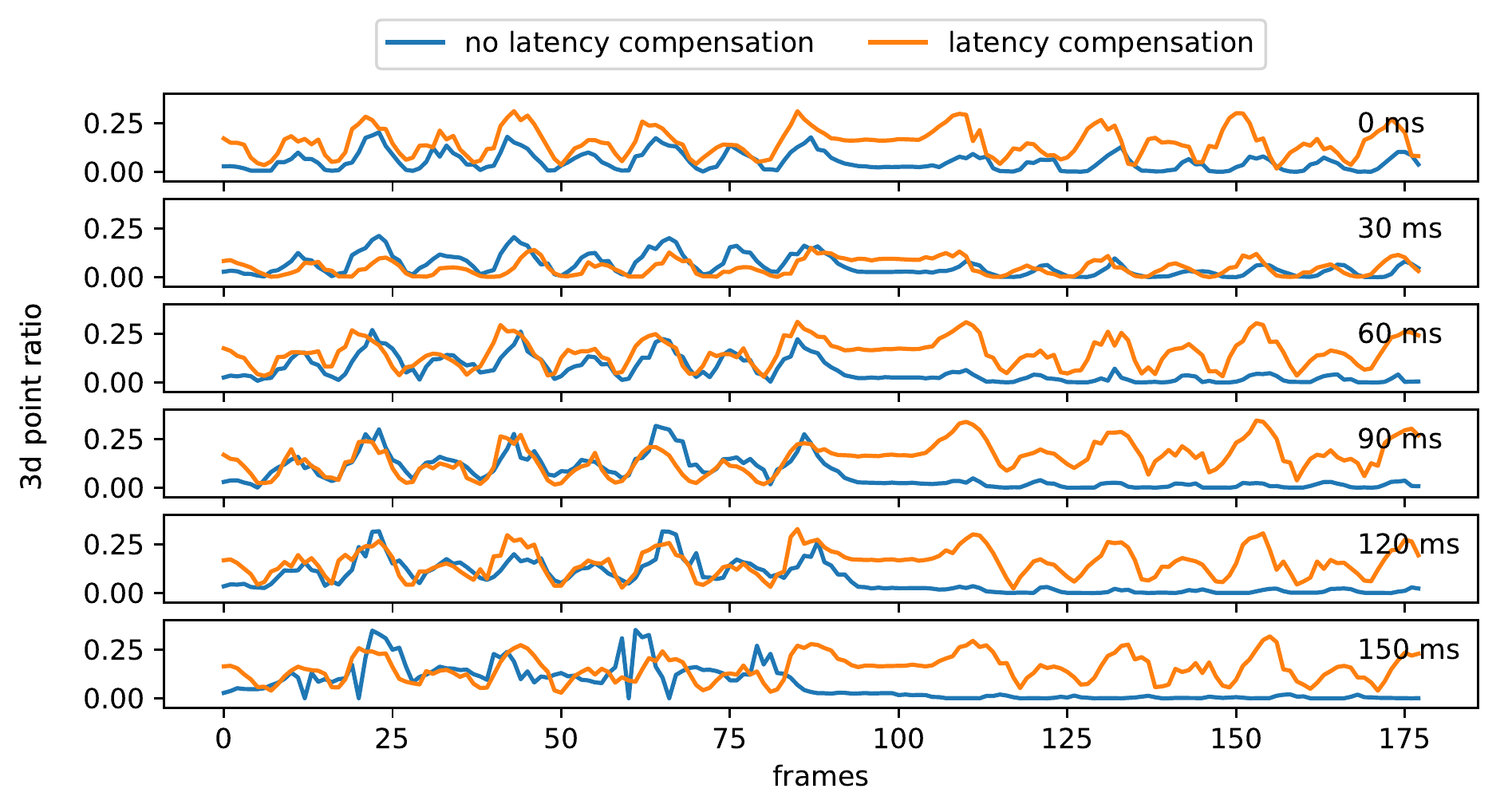}
\end{center}
\vspace{-.3cm}
\caption{Instability of the 3D reconstruction when the latency between two mobiles increases up to 150 ms. The 3D point ratio is defined as the ratio between the number of 3D points and the number of keypoints counted inside the region of interest.}
\label{fig:cp_ratio}
\end{figure}

To illustrate the direct impact of synchronisation delays on the 3D reconstruction, we designed another experiment where our reference object is reconstructed while rotating at an angular velocity of $\omega = 80 deg/s$.
This analysis is performed without using ground-truth information.
We performed various reconstructions of the object by varying the latency between the mobiles.
In one experiment, the object performed two spins: the first spin was 720 degrees counterclockwise, the second spin was 720 degrees clockwise. 
We conducted six experiments in total, where we injected 30ms of communication delay into the WiFi network (using NetEm \cite{NetEm}) between the two mobiles for each experiment. 
Note that if synchronisation triggers are delivered with a delay, the object will appear with a different pose in the two camera frames. 
This affects the computation of the 3D points; matched keypoints will correspond to different 3D locations, and there will be parts of an object that might also be occluded. 
In this first experiment, we could not accurately measure the triangulation error through the ground truth because we did not have direct access to the angular state of the turntable during a spin. 
Therefore, to make 3D reconstructions for each trigger and for each experiment comparable, we quantified the output as being the ratio between the number of 3D points reconstructed and the keypoints visible within the region of interest. 
We compared cases with latency compensation disabled and then enabled.
Fig.~\ref{fig:cp_ratio} shows the variation of the 3D point ratio over time in these two case.
We refer to the left-hand mobile as A and the right-hand as B.
The first 100 frames correspond to 720 degrees of counterclockwise spin. The eight peaks correlate to a face of the box pointing towards both mobiles. 
When latency compensation is disabled, mobile A does not delay the frame capture by $\hat{p}$ ms (Sec.~\ref{sec:frame_capture}) when it sends a synchronisation trigger.
As the induced latency increases, mobile B receives its trigger later and later. 
During this time, the object will have rotated a few degrees in the same the direction as mobile B, resulting in a more favourable viewpoint for keypoint matching and 3D triangulation (i.e.~it is almost seeing an identical view as mobile A). Hence, the 3D point ratio is seen to increase as the induced latency increases when the rotation is counterclockwise. 
However, the computed 3D points will not be calculated correctly and they will also have inaccurate coordinates (we show quantitative evidence of this in the next session). 
Vice-versa, when the rotation is clockwise, mobile A is more likely to capture frames of object regions that are occluded from mobile B's viewpoint as they will have already rotated out of view. 
When our latency compensation algorithm is enabled, 3D reconstructions become symmetric in both spinning directions, illustrating its effectiveness.

\begin{figure}[t]
\begin{center}
\includegraphics[width=1\columnwidth]{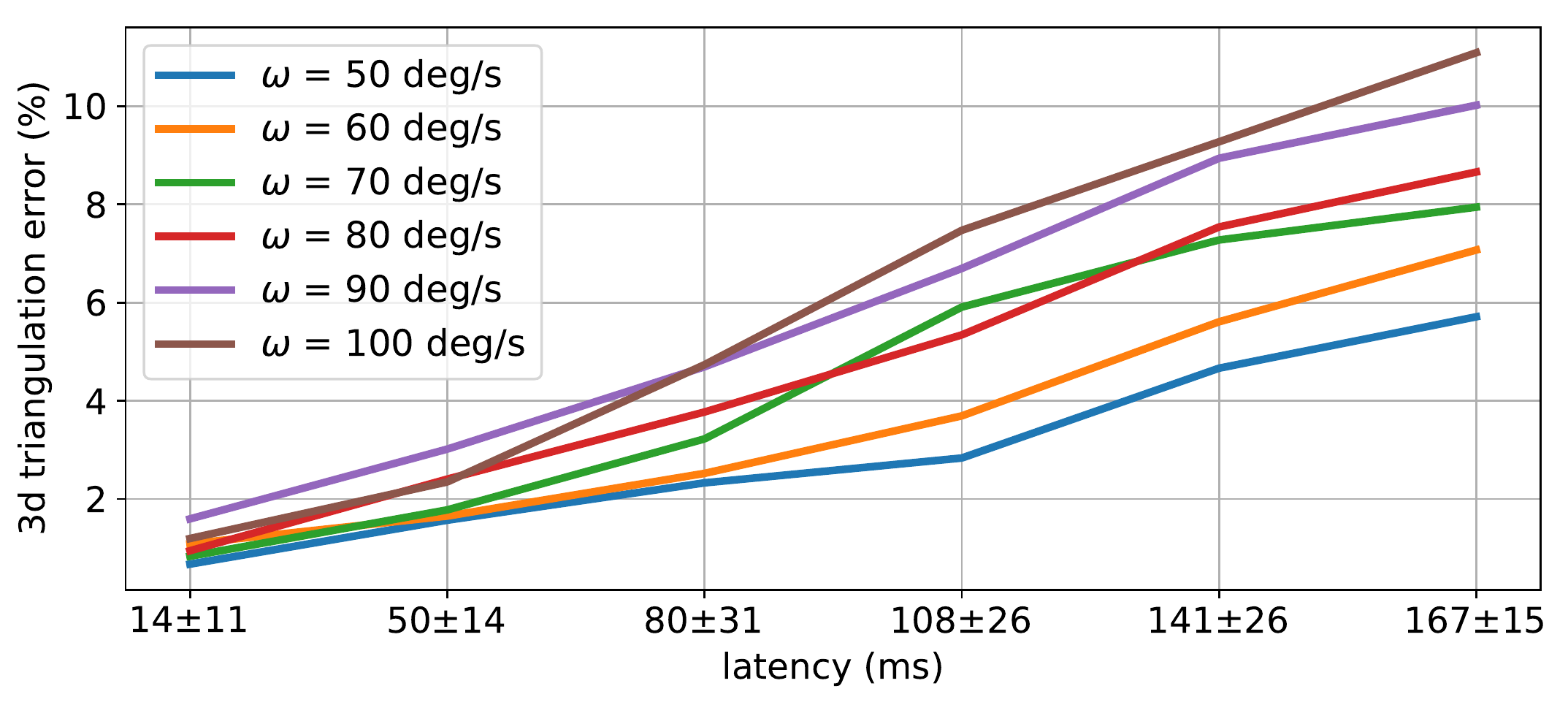}
\end{center}
\vspace{-.2cm}
\caption{3D triangulation error in the case of a counterclockwise-spinning object. The error is computed relative to the distance between the centre of mass of the two cameras and the object, which is 40cm. The camera baseline is fixed at 20cm.}
\label{fig:triang_err_counterclock}
\end{figure}

\begin{figure}[t]
\begin{center}
\includegraphics[width=1\columnwidth]{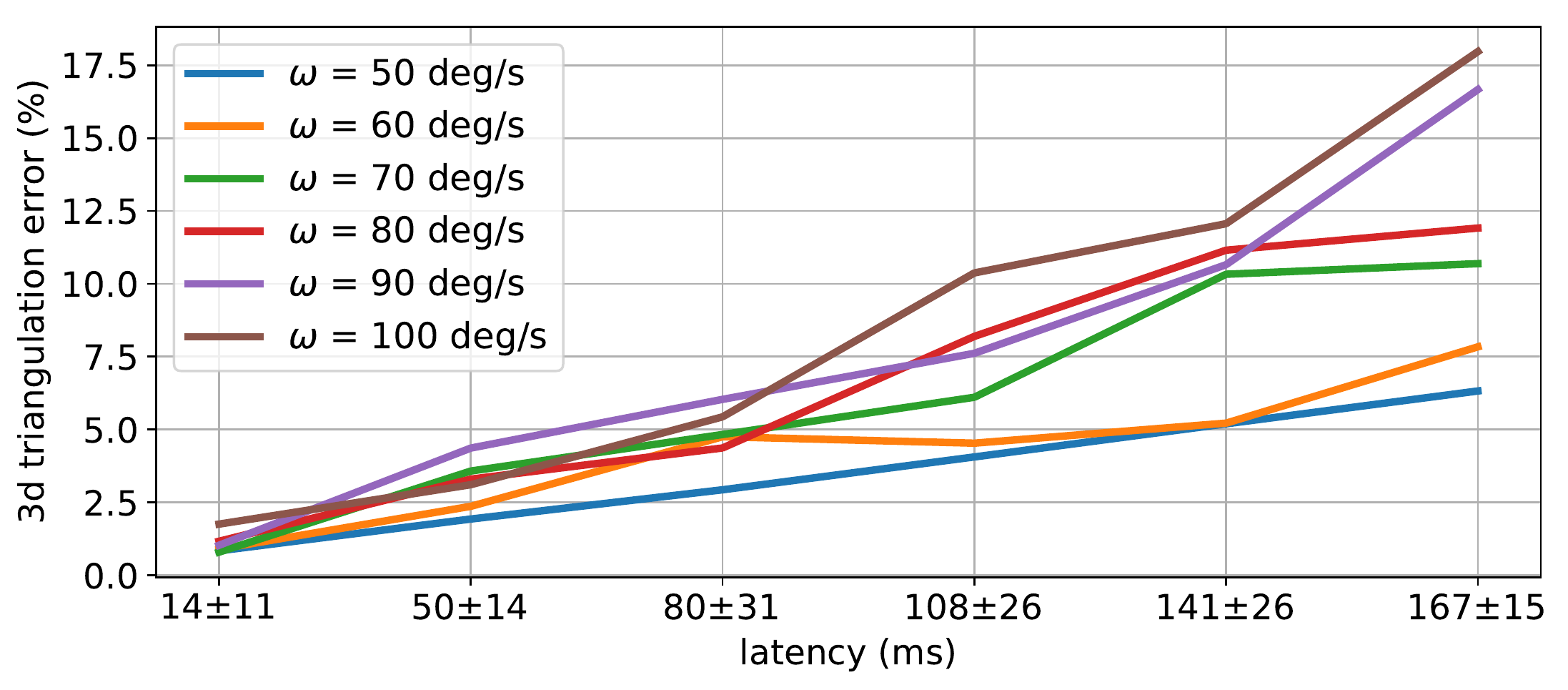}
\end{center}
\vspace{-.2cm}
\caption{3D triangulation error in the case of a clockwise-spinning object. The error is computed relative to the distance between the centre of mass of the two cameras and the object, which is 40cm. The camera baseline is fixed at 20cm.}
\label{fig:triang_err_clock}
\end{figure}

\noindent\textbf{3D reconstruction analysis:}
We quantify the reconstruction accuracy using simulated latency (e.g.~ground truth).
Fig.~\ref{fig:triang_err_counterclock} and \ref{fig:triang_err_clock} show the 3D triangulation error in cases of counterclockwise and clockwise spin, respectively.
From these graphs we see that the triangulation error increases as simulated latency increases. 
This occurs because after keypoints are matched across the two image planes, and, after the keypoints are triangulated in 3D (with an initially-guessed projection matrix), the Bundle Adjustment algorithm in the SfM pipeline tries to optimise the parameters of the projection matrix by minimising re-projection (3D to 2D) error \cite{Schonberger2016}. 
Hence, an erroneous object's pose is captured, thus affecting the estimation of the extrinsic and intrinsic parameters (e.g.~providing different focal-length estimates), and, on the estimation of 3D points (i.e.~highly likely to be estimated somewhere in between the real 3D position of the keypoints of the two frames).
Error-rates in the two cases differ due to the same phenomena illustrated in Fig.~\ref{fig:cp_ratio}. Because synchronisation triggers are generated from the left-hand mobile, which takes snapshots upon generation, the right-hand mobile only takes snapshots when triggers are received. When the object spins clockwise, the larger the delay the more often the object appears with self-occluded parts in the two cameras as it will have rotated more.

\noindent\textbf{End-to-end delay:}
Tab.~\ref{tab:endtoend_delay} shows the end-to-end delay between mobiles, measured as the difference between the times captured from two mobiles through OCR under different communication configurations with the Edge Relay.
As expected, the experiments show evidence that when the Edge Relay is deployed at the edge we can capture frames with the lowest latency.
When the Edge Relay is deployed in the cloud, even through a highly reliable optical-fibre connection, we can see that there is a worsening in the performance due to the extra communication link to AWS.

\begin{table}[t]
\caption{End-to-end delay between mobiles, automatically measured by capturing the time ticked by a stopwatch. Case studies when Edge Relay was deployed in the cloud and at the edge (i.e.~locally). The connection to the cloud was carried out through 4G and a high-quality optical fibre.}
\label{tab:endtoend_delay}
\tabcolsep 7pt
\begin{center}
	\resizebox{\columnwidth}{!}{%
	\begin{tabular}{c|c|c}
	\multicolumn{2}{c|}{Cloud} & \multirow{2}{*}{Edge} \\
	\cline{1-2}
	4G & Optical fibre & \\
	\hline
	$36.36 \pm 25.46$ms & $25.40 \pm 18.68$ms & $20.46 \pm 18.95$ms \\
	\end{tabular}
	}
\end{center}
\end{table}

\begin{figure}[t]
\begin{center}
  \begin{tabular}{@{}c@{}c}
  \vspace{0.3cm}
  \begin{overpic}[width=.47\columnwidth]{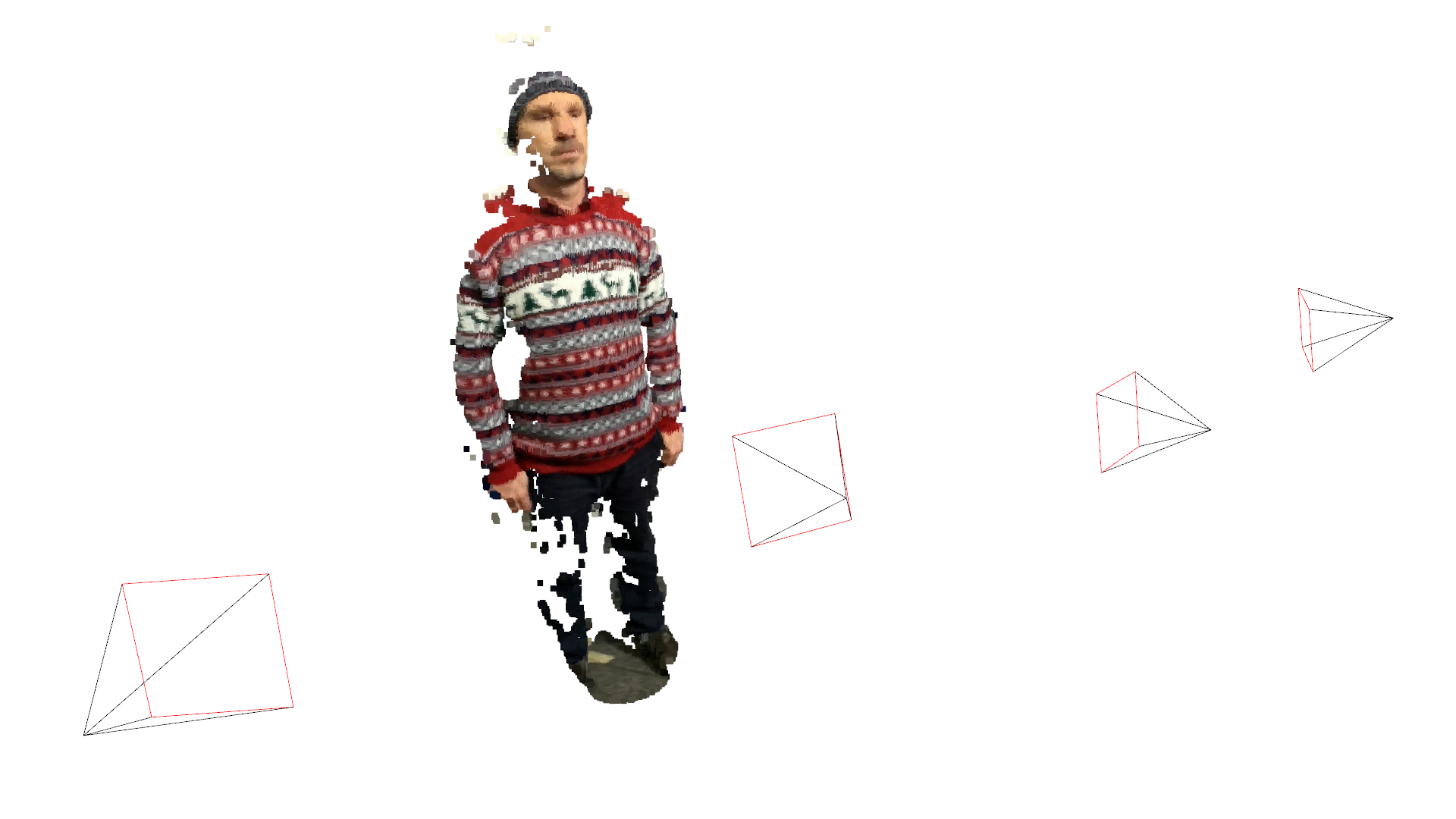}
      \put(15,-6){\color{black}\footnotesize\textbf{(1a) Unity3D Relay}}
    \end{overpic}&
    \begin{overpic}[width=.5\columnwidth]{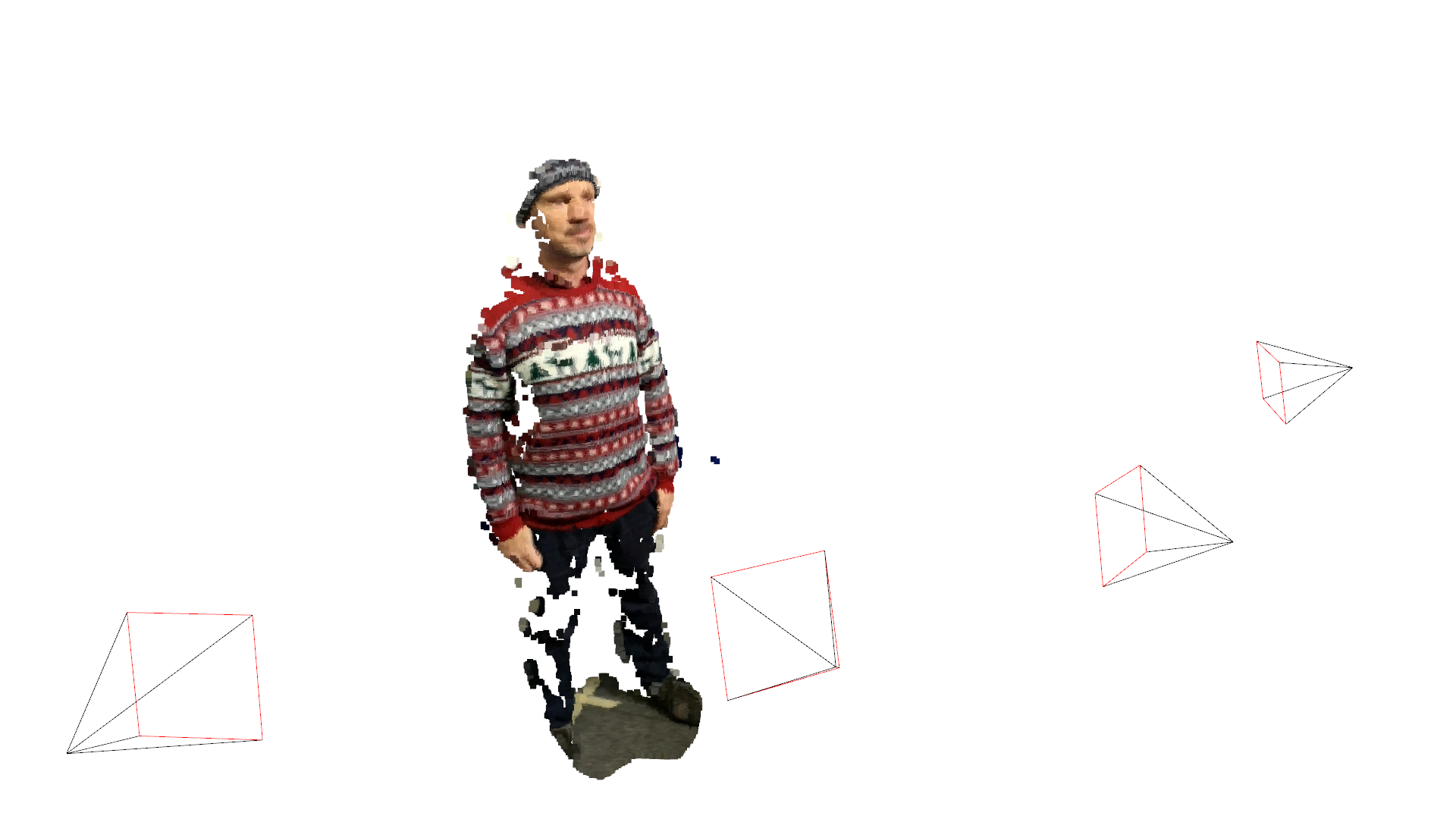}
      \put(15,-6){\color{black}\footnotesize\textbf{(1b) Edge Relay}}
    \end{overpic}\\
    \begin{overpic}[width=.47\columnwidth]{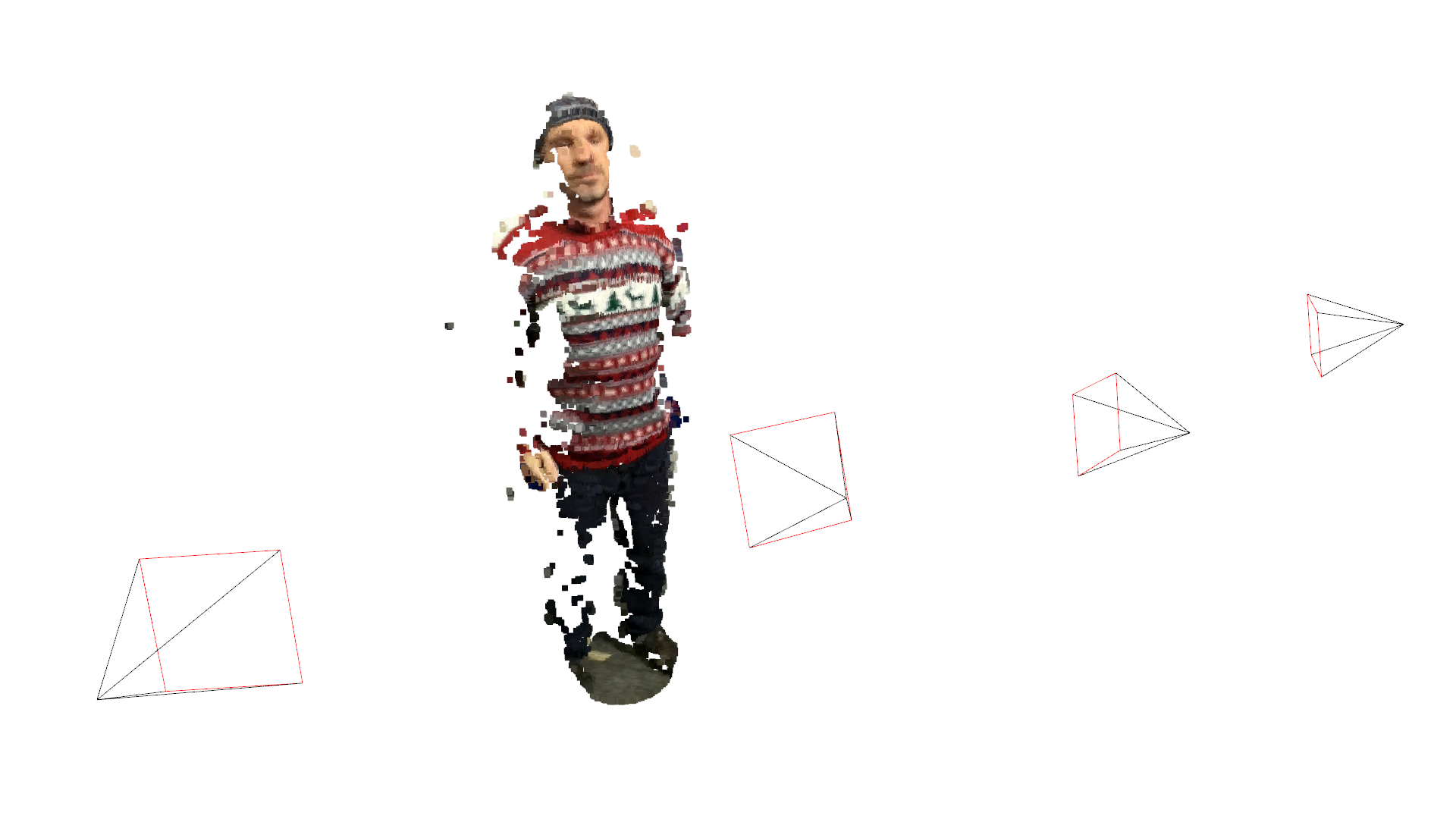}
      \put(15,-6){\color{black}\footnotesize\textbf{(2a) Unity3D Relay}}
    \end{overpic}&
    \begin{overpic}[width=.5\columnwidth]{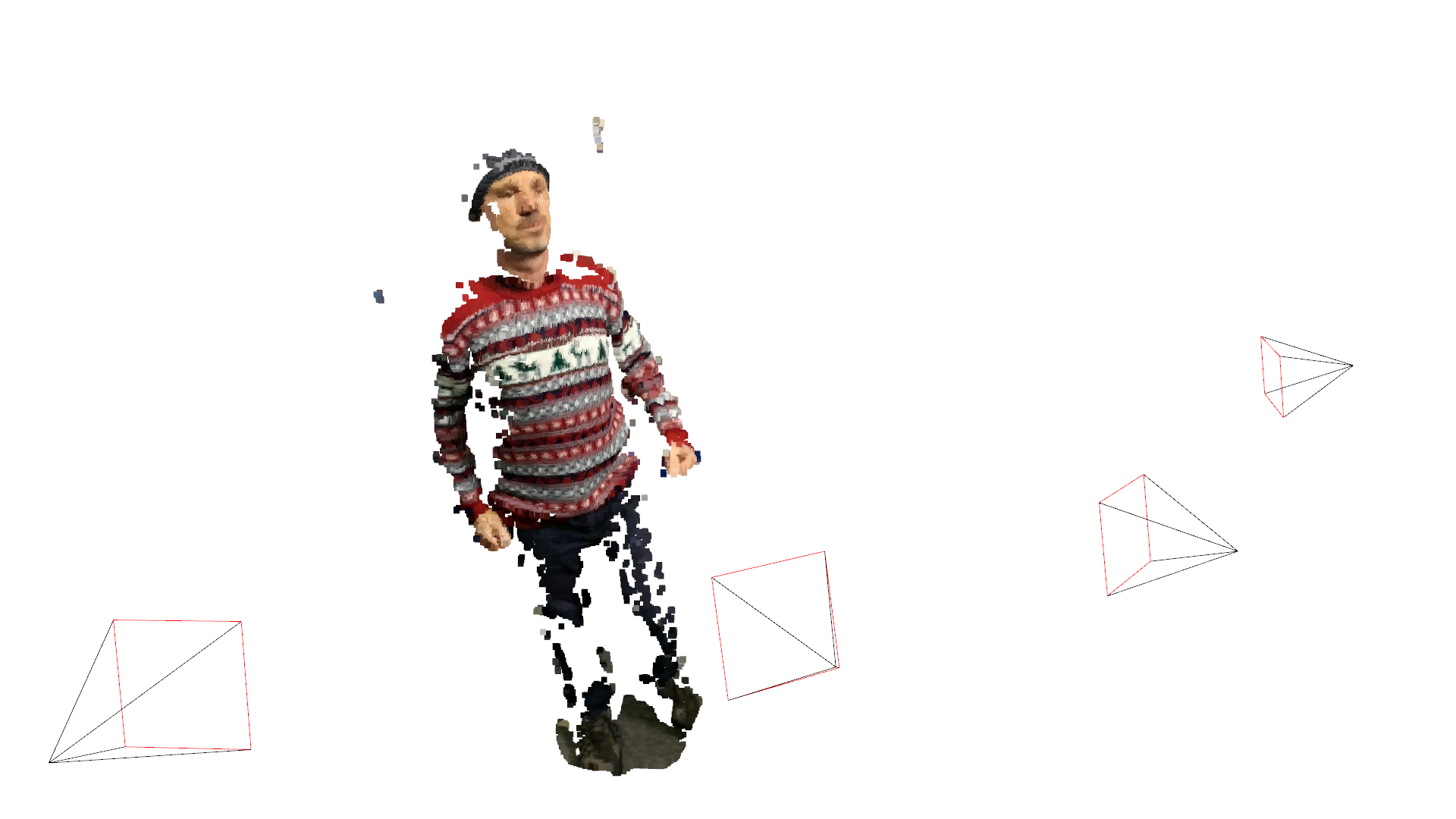}
      \put(15,-6){\color{black}\footnotesize\textbf{(2b) Edge Relay}}
    \end{overpic}\\
  \end{tabular}
\end{center}
\caption{Examples of a moving person reconstructed using (1a,2a) the Unity3D Relay \cite{unity_relay_server} and (1b,2b) our Edge Relay. Case 1: When the object is still we can see that results (i.e.~density of triangulated 3D points) using Unity3D Relay and Edge Relay are comparable. Case 2: When the object moves, synchronisation is key to achieve accurate 3D triangulation, and using the Unity3D Relay leads to sparser reconstructions.}
\label{fig:ex_hlapi_vs_mlapi}
\end{figure}

\noindent\textbf{Qualitative analysis:}
We qualitatively analyse the performance of our approach by comparing the 3D reconstruction over time of a moving person using the standard Unity3D Relay \cite{unity_relay_server} and our Edge Relay. 
We used the AR Anchor to estimate the scale of the point cloud in metric units.
Fig.~\ref{fig:ex_hlapi_vs_mlapi} shows the dense point clouds in two instants of time where (1a,2a) are the outputs with the Unity3D Relay and (1b,2b) with our Edge Relay. We can see that when the object is still (Case 1) the results (i.e.~density of triangulated 3D points) using Unity3D Relay and Edge Relay are comparable. 
Whereas, when the object moves (Case 2), synchronisation is key to achieve accurate 3D triangulation, and using the Unity3D Relay leads to sparser reconstructions.
We quantified the 3D triangulation accuracy by calculating the average reprojection error after Bundle Adjustment.
We measured $0.22 \pm 0.04$ pixels using the Unity3D Relay and $0.20 \pm 0.02$ pixels using our Edge Relay.
This result shows that we could achieve a more accurate reconstruction using the Edge Relay.
During these experiments, we also monitored the percentage of frames successfully received by the Data Manager from all the mobiles.
Given a snapshot trigger, an instant of time that is captured only by a subset of mobiles leads to a partial capture, as only the frames of those mobiles that received the trigger and performed the capture will be transmitted to the Data Manager. 
We name these frames ``partial frames" and, ideally, we would like to achieve zero percent of partial frames. The percentage of partial frames we measured using Unity3D Relay was $41\%$, whereas with our Edge Relay it was $24\%$. Frame loss is because snapshot triggers are transmitted through UDP, that does not acknowledge if packets are received and does not re-send packets in the case of failed reception (unlike with TCP). However UDP is necessary to guarantee the timely delivery of packets. The video of our qualitative analysis can be found at this link \url{https://youtu.be/znoJmovdCgs}. The video illustrates the effect of losing 3D points and frames with the Unity3D Relay.

\section{Conclusions}\label{sec:conclusions}
We proposed a system to move data-relaying from the cloud to the edge, showing that this is key to make frame capture synchronisation more reliable than cloud-based solutions and to enable number-of-users scalability. Our implementation consists of an Edge Relay to handle snapshot triggers used for the capturing of images for FVV production, and a Data Manager to receive capture frames via HTTP requests. 
Synchronisation triggers are generated by a host, rather than by a system timer, to enable a motion-based, adaptive sampling-rate, fostering reduced data throughput.
Although the creation of high-quality FVVs was not the scope of this work, we succeeded to show the benefit of our decentralised data capturing system using a state-of-the-art 3D reconstruction algorithm (i.e.~COLMAP) and by implementing the assessment of end-to-end capture delays though OCR.

Future research directions include the integration of a volumetric 4D reconstruction algorithm that can be executed in real-time on the edge to providing tele-presence functionality together with the integration of temporal filtering of 3D reconstructed points to provide more stable volumetric videos.
We also aim to improve reconstruction accuracy by postprocessing ARCore's pose estimates. By the end of this year we will deploy our system on a 5G network and carry out the first FVV production in uncontrolled environments using off-the-shelf mobiles.

\bibliographystyle{apalike}
{\small
\bibliography{refs}}

\begin{thebibliography}{}

\bibitem[{Amazon Textract}, 2019]{ocr}
{Amazon Textract} (accessed: Nov 2019).
\newblock \url{https://aws.amazon.com/textract/}.

\bibitem[{ARCore}, 2019]{arcore}
{ARCore} (Accessed: Nov 2019).
\newblock \url{https://developers.google.com/ar}.

\bibitem[{ARCore Anchors}, 2019]{anchors}
{ARCore Anchors} (Accessed: Nov 2019).
\newblock
  \url{https://developers.google.com/ar/develop/developer-guides/anchors}.

\bibitem[Bastug et~al., 2017]{Bastug2017}
Bastug, E., Bennis, M., Medard, M., and Debbah, M. (2017).
\newblock {Toward Interconnected Virtual Reality: Opportunities, Challenges,
  and Enablers}.
\newblock {\em IEEE Communications Magazine}, 55(6):110--117.

\bibitem[Belshe et~al., 2015]{RFC7540}
Belshe, M., Peon, R., and Thomson, M. (2015).
\newblock {Hypertext Transfer Protocol Version 2 (HTTP/2)}.
\newblock RFC 7540.

\bibitem[Berners-Lee et~al., 1996]{RFC1945}
Berners-Lee, T., Fielding, R., and Nielsen, H. (1996).
\newblock Hypertext transfer protocol -- http/1.0.
\newblock RFC 1945.

\bibitem[Cadena et~al., 2016]{Cadena2016}
Cadena, C., Carlone, L., Carrillo, H., Latif, Y., Scaramuzza, D., Neira, J.,
  Reid, I., and Leonard, J. (2016).
\newblock {Past, Present, and Future of Simultaneous Localization And Mapping:
  Towards the Robust-Perception Age}.
\newblock {\em IEEE Trans. on Robotics}, 32(6):1309--1332.

\bibitem[Chen et~al., 2016]{Chen2016}
Chen, Y.-H., Balakrishnan, H., Ravindranath, L., and Bahl, P. (2016).
\newblock {GLIMPSE: Continuous, Real-Time Object Recognition on Mobile
  Devices}.
\newblock {\em GetMobile: Mobile Computing and Communications}, 20(1):26--29.

\bibitem[Elbamby et~al., 2018]{Elbamby2018}
Elbamby, M., Perfecto, C., Bennis, M., and Doppler, K. (2018).
\newblock {Toward Low-Latency and Ultra-Reliable Virtual Reality}.
\newblock {\em IEEE Network}, 32(2):78--84.

\bibitem[Guillemaut and Hilton, 2011]{Guillemaut2011}
Guillemaut, J.-Y. and Hilton, A. (2011).
\newblock {Joint Multi-Layer Segmentation and Reconstruction for Free-Viewpoint
  Video Applications}.
\newblock {\em International Journal on Computer Vision}, 93(1):73--100.

\bibitem[Hu et~al., 2016]{Hu2016}
Hu, Y., Niu, D., and Li, Z. (2016).
\newblock {A Geometric Approach to Server Selection for Interactive Video
  Streaming}.
\newblock {\em IEEE Trans. on Multimedia}, 18(5):840--851.

\bibitem[Huang et~al., 2014]{Huang2014}
Huang, C.-H., Boyer, E., Navab, N., and Ilic, S. (2014).
\newblock {Human Shape and Pose Tracking Using Keyframes}.
\newblock In {\em Proc. of IEEE Computer Vision and Pattern Recognition},
  Columbus, US.

\bibitem[Jiang and Liu, 2017]{Jiang2017}
Jiang, D. and Liu, G. (2017).
\newblock {An Overview of 5G Requirements}.
\newblock In Xiang, W., Zheng, K., and Shen, X., editors, {\em 5G Mobile
  Communications}. Springer.

\bibitem[Kim et~al., 2012]{Kim2012}
Kim, H., Guillemaut, J.-Y., Takai, T., Sarim, M., and Hilton, A. (2012).
\newblock {Outdoor Dynamic 3D Scene Reconstruction}.
\newblock {\em IEEE Trans. on Circuits and Systems for Video Technology},
  22(11):1611--1622.

\bibitem[Knapitsch et~al., 2017]{Knapitsch2017}
Knapitsch, A., Park, J., Zhou, Q.-Y., and Koltun, V. (2017).
\newblock {Tanks and Temples: Benchmarking Large-Scale Scene Reconstruction}.
\newblock {\em ACM Transactions on Graphics}, 36(4).

\bibitem[Latimer et~al., 2015]{Latimer2015}
Latimer, R., Holloway, J., Veeraraghavan, A., and Sabharwal, A. (2015).
\newblock {SocialSync: Sub-Frame Synchronization in a Smartphone Camera
  Network}.
\newblock In {\em Proc. of European Conference on Computer Vision Workshops},
  Zurich, CH.

\bibitem[{MLAPI}, 2019]{mlapi}
{MLAPI} (Accessed: Nov 2019).
\newblock \url{https://midlevel.github.io/MLAPI}.

\bibitem[{MLAPI Configuration}, 2019]{mlapi_config}
{MLAPI Configuration} (Accessed: Nov 2019).
\newblock \url{https://github.com/MidLevel/MLAPI.Relay}.

\bibitem[{MLAPI Messaging System}, 2019]{data_messages}
{MLAPI Messaging System} (Accessed: Nov 2019).
\newblock \url{https://mlapi.network/wiki/ways-to-syncronize/}.

\bibitem[Mur-Artal et~al., 2015]{Mur-Artal2015}
Mur-Artal, R., Montiel, J., and Tard\'os, J. (2015).
\newblock {ORB-SLAM}: a versatile and accurate monocular {SLAM} system.
\newblock {\em IEEE Trans. on Robotics}, 31(5):1147--1163.

\bibitem[Mustafa and Hilton, 2017]{Mustafa2017}
Mustafa, A. and Hilton, A. (2017).
\newblock {Semantically Coherent Co-Segmentation and Reconstruction of Dynamic
  Scenes}.
\newblock In {\em Proc. of IEEE Computer Vision and Pattern Recognition},
  Honolulu, US.

\bibitem[{NetEm}, 2019]{NetEm}
{NetEm} (Accessed: Nov 2019).
\newblock \url{http://man7.org/linux/man-pages/man8/tc-netem.8.html}.

\bibitem[Park et~al., 2018]{Park2018}
Park, J., Chou, P., and Hwang, J.-N. (2018).
\newblock Volumetric media streaming for augmented reality.
\newblock In {\em Proc. of IEEE Global Communications Conference}, Abu Dhabi,
  United Arab Emirates.

\bibitem[{Photon}, 2019]{photon}
{Photon} (Accessed: Nov 2019).
\newblock \url{https://www.photonengine.com}.

\bibitem[Poiesi et~al., 2017]{Poiesi2017}
Poiesi, F., Locher, A., Chippendale, P., Nocerino, E., Remondino, F., and Gool,
  L.~V. (2017).
\newblock {Cloud-based Collaborative 3D Reconstruction Using Smartphones}.
\newblock In {\em Proc. of European Conference on Visual Media Production}.

\bibitem[Qiao et~al., 2019]{Qiao2019}
Qiao, X., Ren, P., Dustdar, S., Liu, L., Ma, H., and Chen, J. (2019).
\newblock {Web AR: A Promising Future for Mobile Augmented Reality--State of
  the Art, Challenges, and Insights}.
\newblock {\em Proceedings of the IEEE (Early Access)}.

\bibitem[{QUIC}, 2019]{quic}
{QUIC} (Accessed: Nov 2019).
\newblock \url{https://www.chromium.org/quic}.

\bibitem[Rematas et~al., 2018]{Rematas2018}
Rematas, K., Kemelmacher-Shlizerman, I., Curless, B., and Seitz, S. (2018).
\newblock Soccer on your tabletop.
\newblock In {\em Proc. of IEEE Computer Vision and Pattern Recognition}, Salt
  Lake City, US.

\bibitem[Resch et~al., 2015]{Resch2015}
Resch, B., Lensch, H. P.~A., Wang, O., Pollefeys, M., and Sorkine-Hornung, A.
  (2015).
\newblock Scalable structure from motion for densely sampled videos.
\newblock In {\em Proc. of IEEE Computer Vision and Pattern Recognition},
  Boston, US.

\bibitem[Richardt et~al., 2016]{Richardt2016}
Richardt, C., Kim, H., Valgaerts, L., and Theobalt, C. (2016).
\newblock {Dense Wide-Baseline Scene Flow From Two Handheld Video Cameras}.
\newblock In {\em Proc. of 3D Vision}, Stanford, US.

\bibitem[Schonberger and Frahm, 2016]{Schonberger2016}
Schonberger, J. and Frahm, J.-M. (2016).
\newblock Structure-from-motion revisited.
\newblock In {\em Proc. of IEEE Computer Vision and Pattern Recognition}, Las
  Vegas, USA.

\bibitem[Shi et~al., 2016]{Shi2016}
Shi, W., Cao, J., Zhang, Q., Li, Y., and Xu, L. (2016).
\newblock {Edge Computing: Vision and Challenges}.
\newblock {\em IEEE Internet of Things Journal}, 5(3):637--646.

\bibitem[Shi et~al., 2015]{Shi2015}
Shi, Z., Wang, H., Wei, W., Zheng, X., Zhao, M., and Zhao, J. (2015).
\newblock {A novel individual location recommendation system based on mobile
  augmented reality}.
\newblock In {\em Proc. of IEEE Identification, Information, and Knowledge in
  the Internet of Things}, Beijing, CN.

\bibitem[Slabaugh et~al., 2001]{Slabaugh2001}
Slabaugh, G., Culbertson, B., Malzbender, T., and Schafer, R. (2001).
\newblock {A Survey of Methods for Volumetric Scene Reconstruction from
  Photographs}.
\newblock In {\em International Workshop on Volume Graphics}, New York, US.

\bibitem[Soret et~al., 2014]{Soret2014}
Soret, B., Mogensen, P., Pedersen, K., and Aguayo-Torres, M. (2014).
\newblock {Fundamental tradeoffs among reliability, latency and throughput in
  cellular networks}.
\newblock In {\em Proc. of IEEE Globecom Workshops}, Austin, US.

\bibitem[Sukhmani et~al., 2018]{Sukhmani2018}
Sukhmani, S., Sadeghi, M., Erol-Kantarci, M., and Saddik, A.~E. (2018).
\newblock {Edge Caching and Computing in 5G for Mobile AR/VR and Tactile
  Internet}.
\newblock {\em IEEE Multimedia (Early Access)}.

\bibitem[{Unity3D HLAPI}, 2019]{hlapi}
{Unity3D HLAPI} (accessed: Nov 2019).
\newblock \url{https://docs.unity3d.com/Manual/UNetUsingHLAPI.html}.

\bibitem[{Unity3D MatchMaking}, 2019]{matchmaking}
{Unity3D MatchMaking} (Accessed: Nov 2019).
\newblock
  \url{https://docs.unity3d.com/520/Documentation/Manual/UNetMatchMaker.html}.

\bibitem[{Unity3D Multiplayer Service}, 2019]{unity_relay_server}
{Unity3D Multiplayer Service} (Accessed: Nov 2019).
\newblock \url{https://unity3d.com/unity/features/multiplayer}.

\bibitem[{Unity3D Network Manager}, 2019]{unity_match_host}
{Unity3D Network Manager} (Accessed: Nov 2019).
\newblock
  \url{https://docs.unity3d.com/ScriptReference/Networking.NetworkManager-matchHost.html}.

\bibitem[Vo et~al., 2016]{Vo2016}
Vo, M., Narasimhan, S., and Sheikh, Y. (2016).
\newblock {Spatiotemporal Bundle Adjustment for Dynamic 3D Reconstruction}.
\newblock In {\em Proc. of IEEE Computer Vision and Pattern Recognition}, Las
  Vegas, US.

\bibitem[Wang et~al., 2015]{Wang2015}
Wang, Y., Wang, J., and Chang, S.-F. (2015).
\newblock {CamSwarm: Instantaneous Smartphone Camera Arrays for Collaborative
  Photography}.
\newblock {\em arXiv:1507.01148}.

\bibitem[{Wu} et~al., 2008]{Wu2008}
{Wu}, W., {Yang}, Z., {Jin}, D., and {Nahrstedt}, K. (2008).
\newblock Implementing a distributed tele-immersive system.
\newblock In {\em 2008 Tenth IEEE International Symposium on Multimedia}, pages
  477--484.

\bibitem[Yahyavi and Kemme, 2013]{Yahyavi2013}
Yahyavi, A. and Kemme, B. (2013).
\newblock Peer-to-peer architectures for massively multiplayer online games: A
  survey.
\newblock {\em ACM Comput. Surv.}, 46(1):1--51.

\bibitem[Zhang et~al., 2018]{Zhang2018}
Zhang, W., Han, B., and Hui, P. (2018).
\newblock {Jaguar: Low Latency Mobile Augmented Reality with Flexible
  Tracking}.
\newblock In {\em Proc. of ACM International Conference on Multimedia}, Seoul,
  KR.

\bibitem[Zou and Tan, 2013]{Zou2013}
Zou, D. and Tan, P. (2013).
\newblock {COSLAM: Collaborative visual slam in dynamic environments}.
\newblock {\em IEEE Trans. on Pattern Analysis and Machine Intelligence},
  35(2):354--366.

\end{thebibliography}

\end{document}